# Time series of Internet AS-level topology graphs: four patterns and one model

Liandong Liu and Ke Xu
State Key Lab. of Software Develop Environment
Beihang University
Beijing, China
{lld, kexu}@nlsde.buaa.edu.cn

*Abstract*—Researchers have proposed a variety of Internet topology models. However almost all of them focus on generating one graph based on one single static source graph. On the other hand, Internet topology is evolving over time continuously with the addition and deletion of nodes and edges. If a model is based on all the topologies in the past, instead of one of them, it will be more accurate and closer to the real world topology. In this paper, we study the Internet As-level topology time-series from two different sources and find that both of them obey four same dynamic graph patterns. Then we propose a mode that can infer the topology in the future based on all the topologies in the past. Through theoretical and experimental analysis, we prove the topology that our model generates can match both the static and dynamic graph patterns. In addition, the parameters in the model are meaningful. Finally, we theoretically and experimentally prove that these parameters are directly related to some important graph characteristics.

## I. INTRODUCTION

Internet topology analysis and modeling has attracted broad attention recently because it could improve our understanding of routing, performance of applications, protocols ,network attack and so on [1, 3, 7-9, 14, 22, 26, 29, 32, 25]. Some studies have been done in this area, and there have been a lot of remarkable results. The power law degree distribution is the most famous one among them. But almost all the studies focus on finding and reproducing the static graph patterns. In this paper, we refer the topology properties of Internet which are learned from one single graph by static graph patterns, including distance distribution, clustering coefficient, betweenness and so on [3, 19, 22].

On the other hand, Internet is a dynamic system which is evolving over time. So some other problems are involved. For example, does the number of edges grow linearly with the number of nodes? Are all the new nodes the same and have the same new edges? Where a new edge can be added probably? All the questions cannot be answered based on one single graph. A time series of topology graphs must be involved in. We will answer all the questions above in this paper.

We study the time series of Internet AS-level topology graphs extracted from two different projects [35, 39]. The following four patterns are found in both of them.

(1) The number of edges grows super-linearly with the number of nodes. Specifically, it follows a power-law pattern,

$$|E_i| \propto |N_i|^\alpha \qquad (1)$$

where $|E_i|$ is the number of edges and $|N_i|$ is the number of nodes at time $i$.

(2) With a low probability, an edge between two new nodes is added.

(3) The new nodes have different initial degree (new edges). Furthermore, the initial degree distribution follows a power-law pattern.

(4) The shorter the distance between two nodes, the higher the probability they will become neighbors at next time point.

These four patterns can give answers to the questions above. Different from static graph patterns, all of them must be learned from a time series of topology graphs. In this paper, we refer this kind of patterns by dynamic graph patterns.

Compared with static graph patterns, dynamic graph patterns were studied by fewer researchers, and only a few models that match dynamic graph patterns were developed [16, 25]. We propose a new model in this paper, which takes as input the time series of graphs in the past, the number of new nodes $\Delta N$, the stable factor of nodes $a_n$, the stable factor of edges $a_e$, and the clustering factor $p$. After three steps (preparation, initialization and generation), it outputs the topology graph at next time point. It can be easily extended to generate a time series of graphs. The graph which the model generates can match both the static graph patterns and the four dynamic graph patterns we find. The model is validated by our experiments.

In addition, the three parameters ($a_v$, $a_e$ and $p$) our model uses are of great significance. It has a direct relationship with some characteristics of the output graph. We will discuss this below with both theoretical and experimental methods.

The rest of the paper is organized as follows. Section 2 introduces the background about topology model and data that we use in this paper. Section 3 gives a number of definitions and related symbols .Section 4 describes graph patterns we focus on, including both static graph patterns and the four dynamic patterns we find. Section 5 introduces the new model in detail. Section 6 validates the model through some experiments. Section 7 discusses the three parameters used in the model. Section 8 and Section 9 discuss the related work and conclude the paper respectively.

## II. BACKGROUND

In this section, we overview some related work, introduce the data source we use and mention several well-know topology models at presents.

### A. Data

The Internet consists of thousands of connected networks called Autonomous Systems (AS) and the Border Gateway Protocol (BGP [33]) is used to exchange reachable information among Autonomous Systems. The entire Internet can be viewed as an AS-level topology graph where each AS is a node and the BGP peering relationship between two ASes is an edge.

We can get the AS-level topology graph from three types of sources: traceroute measurements, BGP data, and Internet registries.

The Skitter [35], DIMES [38], and iPlane[37] are the projects that collect the data using the traceroute-based measurement. They place monitors in the global Internet to periodically traceroute thousands of destination IP address, convert route paths to AS paths, and then construct the entire Internet As-level topology graph. They differ in the numbers of monitors, locations of monitor, probing frequency, and the list of destination IP address. The oldest and the most famous one among them is the Skitter developed by CAIDA, which has the AS-level data from January, 2000 to February, 2008.

Using placing peers (or vantage points), Route Views [34] and RIPE [36] can collect the BGP routing table and update data and then infer the AS-level topology graph. Internet registries from Regional Internet Registries (RIR) or Internet Routing Registries (IRR) are also sources to extract the AS-level topology graph.

Combining data from more than one source may be a way to construct more accurate AS-level graph. The researchers from UCLA [31, 39] make effort in this direction. They extract the most complete AS-level topology from as many inter-domain routing sources as they can, including Route Views, RIPE, route servers, looking glasses, Internet registries, and routing update. Since they update the topology daily, they have data from January, 2004 to now. In order to get time series of topology graphs, we prefer the data collected by Skitter and UCLA. In the following, we will refer them by skitter and ucla respectively.

### B. Topology Model

Modeling Internet growth is of great importance both for understanding the current Internet and predicting its future. Many researchers make effort in this way [1, 2, 5, 7, 8, 12, 15, 17, 20, 30, 32]. The classical ER random graph model is one of the earliest topology models. The BA model [2] is the most well-known model, which is based on the idea "the rich get richer" and can reproduce the observed power-law degree distribution. Some studies extended BA model, such as EBA [1], which involves adding and rewiring edges. A survey about topology model can be found in [4].

Though they differ in details, they all take one graph as input and aim to reproduce some static graph patterns For example, the BA model and its extensions are proposed to match the power-law degree distribution. Our model takes a time series of graphs as input and aims to reproduces both the static and dynamic graph patterns.

## III. PROBLEM DEFINITION

With the changes of nodes and edges, Internet topology evolves continuously. Most of the models at present are based on only one single graph. Since some information may be lost and dynamic graph patterns cannot be reproduced. If a graph in the future is inferred based on a time series of Internet topology graphs in the past, it may be closer to the one in the future, containing more information. Inferring graph based on a time series of graphs (IGBTSG) is a new problem. We will make some efforts to solve it in this paper.

In this section, we give a definition of this problem and introduce a number of related symbols. These notions will be used in developing our model.

### A. Definition of IGBTSG

Given $G_1, G_2, \cdots, G_n$ which are the Internet topology graph at the time of $t_1, t_2, \cdots, t_n$, the IGBTSG problem is to infer $G_{n+1}$, the Internet topology graph at the time of $t_{n+1}$. $G_i$ ($G_i = (V_i, E_i)$) is the Internet topology graph at the time of $t_i$, where $V_i$ represents the set of nodes and $E_i$ represents the set of edges.

In addition, if a model can solve this problem, it must have the following properties.

- If a node or an edge appears in $G_1, G_2, \cdots, G_n$, it may appear in $G_{n+1}$ too.

- The similarity between $G_{n+1}$ and $G_n$ is the highest. The smaller $t$ is, the lower the similarity between $G_{n+1}$ and $G_t$ is.

- $G_{n+1}$ must not only obey the same static graph patterns as $G_1, G_2, \cdots, G_n$, but also the same dynamic graph patterns.

- The parameters which the model employs should have specific meanings. When the parameters are adjusted, the changes of the output graph are predictable.

In order to describe our model more clearly, we introduce some definitions and symbols used later.

## B. Definition of total graph

A graph $G_i^{Total} = \left(V_i^{Total}, E_i^{Total}\right)$ is a union of $G_1, G_2, \cdots, G_i$, where $V_i^{Total} = V_1 \cup V_2 \cup \cdots \cup V_i$ is the union of the node sets and $E_i^{Total} = E_1 \cup E_2 \cup \cdots \cup E_i$ is the union of the edge sets.

Each node $v \in V_i^{Total}$ has a sequence of states (appearance or disappearance) $\{s_0^v, s_1^v, s_2^v, \cdots, s_i^v\}$, where

$$s_k^v = \begin{cases} 1 & v \in V_k \\ 0 & v \notin V_k \end{cases}, (1 \leq k \leq i). \quad (2)$$

And for convenience, we define an initial state $s_0^v$. $s_0^v$ is equal to 0 for all nodes.

Similarly, each edge $e \in E_i^{Total}$ has a sequence of states (appearance or disappearance) $\{s_0^e, s_1^e, s_2^e, \cdots, s_i^e\}$, where

$$s_k^e = \begin{cases} 1 & e \in E_k \\ 0 & e \notin E_k \end{cases}, (1 \leq k \leq i). \quad (3)$$

We define an initial state $s_0^e$ too. $s_0^e$ is equal to 0 for all edges.

$G_0^{Total}$ is an empty graph.

## C. Definition of new node and old node

If a node $v \in V_i^{Total}$ and $v \notin V_{i-1}^{Total}$, $v$ is called a new node at the time of $t_i$ and $V_i^{new}$ represents the set of all new nodes at the time of $t_i$. The degree (number of neighbors) of $v$ in $G_i$ is called initial degree of $v$. Similarly, if a node $v \in V_i^{Total}$ and $v \in V_{i-1}^{Total}$, $v$ is called an old node at the time of $t_i$ and $V_i^{old}$ represents the set of all old nodes at the time of $t_i$.

## D. Definition of new edge

If an edge $e \in E_i^{Total}$ and $e \notin E_{i-1}^{Total}$, $e$ is called a new edge at the time of $t_i$. Because the two nodes $v$ and $u$ which $e$ connects belong to different types, we divide all the new edges into three parts $E_i^{new-new}$, $E_i^{new-old}$ and $E_i^{old-old}$. if $v$ and $u$ are both new nodes, $e \in E_i^{new-new}$, if $v$ and $u$ are both old nodes, $e \in E_i^{old-old}$, otherwise, $e \in E_i^{new-old}$.

## IV. GRAPH PATTERN

The graph the model generates must match both static and dynamic graph patterns of the source graph. In this section, we give a list of graph patterns which our model should fit.

### A. Static graph patterns

Many studies have been done to analyze Internet topology. A lot of static graph patterns have been found, studied and validated. We outline some important ones among them.

(a) The node degree distribution is the probability that a randomly selected node is $k$-degree: $P(k) = \frac{n(k)}{n}$, where n is the number of nodes and $n(k)$ is the number of nodes whose degree is $k$. Many studies found the node degree distribution in Internet as-level topology graph is power law, i.e. $P(k) \propto k^{-\gamma}$, where $\gamma$ is a positive exponent.

(b) The distance distribution $d(x)$ is the number of pairs of nodes at a distance $x$, divided by the total number of node pairs $C_n^2 = \frac{n(n-1)}{2}$ (self-pairs excluded).

(c) The local clustering $C(k)$ is the ratio of this average number of edges between the neighbors of k-degree nodes to the maximum number of such edges $C(k) = \frac{\overline{m}(k)}{C_k^2}$, where $\overline{m}(k)$ is the average number of edges between the neighbor of k-degree nodes, and $C_k^2 = \frac{k(k-1)}{2}$ is the maximum number of such edges. The mean local clustering $\overline{C} = \sum C(k)P(k)$

Because of the limitation of the paper length, other static graph patterns such as the spectrum of graph, betweenness, and assortative coefficient are not discussed in this paper. More information about topology characteristics can be found in [19, 22].

### B. Dynamic graph patterns

The studies about dynamic graph patterns are much fewer than those about static graph patterns. We study the temporal Internet topology graph, by observing datasets from skitter and UCLA. We found four dynamic graph patterns.

(1) The number of edges grows super-linearly with the number of nodes. Specifically, it follows a power-law pattern,

$$|E_i| = A|N_i|^\alpha \quad (4)$$

where $|E_i|$ is the number of edges and $|N_i|$ is the number nodes at time $i$. In Figure 1 (a) and (b), we can see the exponent $\alpha$ is between 1 and 2. The average degree of the graph will get greater over time with the increase of $|N_i|$. So this pattern is called densification power law in [16]. In addition, we found the number of edges and the number of nodes in total graph also follows a similar power-law pattern (Figure 1(c)),

$$|E_i^{Total}| = A'|N_i^{Total}|^{\alpha'} \quad (5)$$

(2) With a low probability, an edge between two new nodes is added. In Figure 1(d), we can see for all the graphs, the percentage of new-new edges in all the new edges are very low (< 0.005).

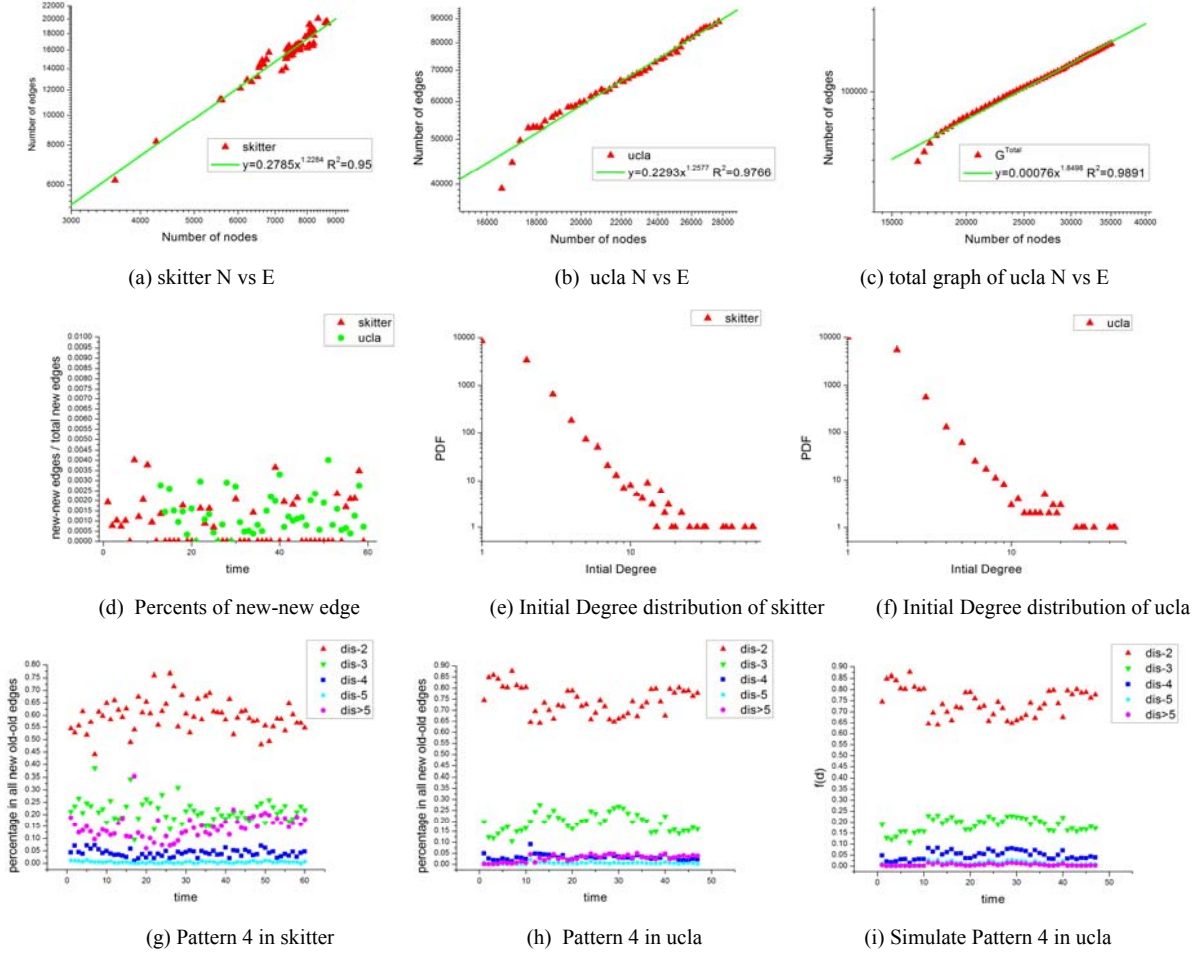

Figure 1. Dynamic Graph Patterns

This suggests the new nodes are added to the graph independently, instead of forming sub-graphs with other new nodes.

(3) The new nodes have different initial degrees. In addition, the initial degree distribution of all the new nodes in the past $P_{initial}(k)$ follows a power-law pattern. This can be seen in Figure 1(e) and (f). The initial degree distribution at a specific time point can be seen as a sample from $P_{initial}(k)$.

(4) The shorter the distance between the two nodes, the higher the probability they will become neighbors at next time point.

We plot the percentage of new old-old edges with different distances between the two endpoint in past in Figure 1 (g) and (h). We can see the proportion of "dis-2" is the largest. If we denote $f(d)$ as the percentage of adding new old-old edges, where d is the distance between the two endpoints. $f(d)$ is expected to

$$\forall d \geq 2, f(d) = p(1-p)^{d-2} \quad (6)$$

where $p$ is the percentage of adding new old-old edges, and the distance between their two endpoint is 2. With (6), we simulate this pattern in ucla. Looking at Figure 1(h) and (i), the pattern that simulated by $f(d)$ is very close to the real one.

## V. PROPOSED MODEL

### A. Overview

In the last section, we have learned that there are four dynamic graph patterns in the Internet AS-level topology graph. In this section, we would like to propose a model that can reproduce both static and dynamic graph patterns. Taking the topology from time *0* to time *n* as input, the model can infer the topology at time *n+1*.

Our model takes as input the graph series in the past $G_1, G_2, \cdots, G_n$, the number of new nodes $\Delta N$, the stable factor of nodes $a_n$, the stable factor of edges $a_e$, and the clustering factor $p$. In the rest of this section, we will explain them in detail. The process can be mainly divided into three steps: preparation, initialization, generation. In the preparation step, the model will build the $G_n^{Total}$ based on $G_1, G_2, \cdots, G_n$, estimate the number of new edges $\Delta E$, and prepare the initial degree distribution of new nodes. In the initialization step, the model will create an initial output graph based on $G_n^{Total}$, which has no new edges and nodes. In the generation step, the model will finish generating through adding new nodes and edges into the initial graph.

In the rest of this section, we provide additional details on how these operations are performed.

*B. Preparation*

In the preparation step, the model mainly performs three operations.

First of all, it will build the $G_n^{Total}$. As shown in Section 3, $G_n^{Total}$ is a union of $G_1, G_2, \cdots, G_n$. The most important difference between total graph and an ordinary graph is both the edges and nodes in $G_n^{Total}$ have a sequence of states which can show whether they appeared in the past. In the building process, we also get some secondary products such as the initial degree distribution of all the new nodes $P_{initial}(k)$ before the time $n$.

Secondly, Equation (7) are derived from (5)

$$\Delta E = A' \left| N_n^{Total} \right|^{\alpha'-1} \Delta N \qquad (7)$$

If we would like to make the number of nodes and the number of edges in $G_{n+1}^{Total}$ and $G_{n+1}$ obey the same pattern as before, the model must add $A' \left| N_n^{Total} \right|^{\alpha'-1} \Delta N$ edges while adding $\Delta N$ nodes.

Finally, in order to make the initial degree distribution of nodes in $G_{n+1}^{Total}$ obey the power law, we use the following procedure to prepare $n(k)$ (the number of nodes with $k$ initial degree in $G_{n+1}$),

(1) Find the threshold $k_s$ defined as the highest initial degree value that satisfy the condition:

$$\forall k \leq k_s, P_{initial}(k) > 0, P_{initial}(k) > P_{initial}(k+1)$$

For $k \leq k_s$, the initial degree distribution PDF follows a power law approximately. For $k > k_s$, the distribution is in the heavy-tailed range.

(2) For $k \leq k_s$, the number of nodes with k initial degree $n(k) = \Delta N \times P_{initial}(k)$. Since $\Delta N \times P_{initial}(k)$ may be not an integer, we set $n(k) = \lfloor \Delta N \times P_{initial}(k) + 0.5 \rfloor$ in practice.

(3) The initial degree has been set for $\sum_{i=1}^{k_s} n(i)$ nodes, for the rest nodes ( the number is $\Delta N - \sum_{i=1}^{k_s} n(i)$ ), their initial degree will be chosen randomly between $k_s + 1$ and $\Delta N \frac{K_{max}}{\Delta N_{max}}$, where $K_{max}$ is the maximum initial degree and $\Delta N_{max}$ is the maximum number of new nodes in $G_1, G_2, \cdots, G_n$.

After the three procedures, the model finishes preparing $n(k)$.

*C. Initialization*

In the initialization step, the model will create an initial output graph. The biggest difference between the initial output graph and the final one is that all the nodes and edges in the former are from $G_n^{Total}$. After adding new nodes and edges into the initial one in the next step, the model will finish generating the final output graph. The method used in this step is based on the idea below:

- If a node (or an edge) appears in $G_1, G_2, \cdots, G_n$, it may appear in $G_{n+1}$ too.

- The similarity between $G_{n+1}$ and $G_n$ is the highest. The smaller $t$ is, the lower the similarity between $G_{n+1}$ and $G_t$ is. We can say the similarity between $G_{n+1}$ and $G_t$ is high when the state of nodes and edges in $G_{n+1}$ is same as $G_t$.

First we involve two parameters, the stable factor of nodes $a_n$ and the stable factor of edges $a_e$.

For each node $v \in V_n^{Total}$, and its sequence of states $\{s_0^v, s_1^v, s_2^v, \cdots, s_n^v\}$, we define the probability $p_i^V$ that $s_{n+1}^v$ is equal to $s_i^v$ as:

$$p_i^V = \begin{cases} a_n (1-a_n)^{n-i} & 1 \leq i \leq n \\ (1-a_n)^n & i = 0 \end{cases} \qquad (8)$$

We can know from the definition, with a probability $a_n$, the state of a node in $G_{n+1}$ is the same as $G_n$. Because the value $a_n$ is between 0 and 1, the probability will decline with the decrease of $i$. In addition, we can validate that the sum of $p_i^V$ is equal to 1.

Furthermore, we define the mean value of $v$'s sequence of states as its expected state in $G_{n+1}$

$$E(s^v) = \sum_{i=0}^{n} p_i^v s_i^v \qquad (9)$$

For each edge $e \in E_n^{Total}$ and its sequence of states $\{s_0^e, s_1^e, s_2^e, \cdots, s_n^e\}$, we similarly define the probability that $s_{n+1}^e$ is equal to $s_i^e$ as :

$$p_i^E = \begin{cases} a_e (1-a_e)^{n-i} & 1 \leq i \leq n \\ (1-a_e)^n & i = 0 \end{cases} \qquad (10)$$

and we define the mean value of *e*'s sequence of states as its expected state in $G_{n+1}$

$$E(s^e) = \sum_{i=0}^{n} p_i^E s_i^e \qquad (11)$$

Now we describe the initialization step. First, for each node in $G_n^{Total}$, the model will generate a random value (between 0 and 1). If the random value is smaller than the node's expected state, this node will be added into the initial graph, else not. Then, for each edge in $G_n^{Total}$, the model will run similarly, generate a random value, and add the edge into the initial graph when the random value is smaller than its expected state. If a node is not added into the initial graph, all the edge connecting to it is not added into the initial graph either.

### D. Generation

In the step, the model will add new nodes and edges into the initial graph created in the last step, and then finish the generation of $G_{n+1}$. The model will perform two operations: adding new old-old edges and adding new nodes. Both of them requires a sub-procedure, given a source node, how to select the target node and create an edge between them. So we introduce this sub-procedure first. We call it Distance Guided Attachment in this paper.

**Distance Guided attachment**

Given a source node, the model uses the following operations to select another node and create an edge between them.

The model divides the nodes in initial graph into sets based the distance from the source node to them. We denote the nodes set whose distance to source node is equal to *d* by *d-set* ($d \geq 2$). we denote the nodes which the source node can not arrive by $\infty$-*set*.

With probability $f(d) = p(1-p)^{d-2}$, the model selects the target node from *d-set*. In order to make the output graph obey the power law degree distribution, the model also uses a "rich get richer" attachment process. The target node is selected with probability $\dfrac{k_i+1}{\sum_{j \in d-set}(k_j+1)}$ in *d-set*, where $k_j$ denote the degree of node *j*.

With probability $1-\sum_{d=2}^{d_{max}} p(1-p)^{d-2}$, the model selects the target node from $\infty$-*set*, and where $d_{max}$ is the longest distance from source node to others. Certainly the model uses a "rich get richer" attachment process when selecting a node from $\infty$-*set*.

From the procedure, we can see when the node is closer to the source node (*d* is smaller), with a higher probability an edge between them is added. This make the output graph match the pattern 4 in Section 4.

Now we introduce the generation step in detail.

First the model adds all the new nodes into the initial graph.

Then the model prepares an empty source node list. For each new node, model adds it into source node list *k* times (*k* is the initial degree of this node).

From (7), the total number of new edges is gotten. From the initial degree distribution of new nodes, the number of new-old edges is gotten. The former minus the latter is the number of old-old edges $|E_{n+1}^{old-old}|$. The model selects $|E_{n+1}^{old-old}|$ old nodes from the initial graph randomly and then adds them into the source node list.

Finally the model randomly shuffles the source node list first. For each node in the list, with the Distance Guided Attachment method, the model selects a target node and creates an edge between them. If the source node is an old node and the target node is a new node, the model will select the target node again until the target node is an old node. In addition, because compared with old nodes the new nodes is very little, the edges between new nodes may exit but must be very little.

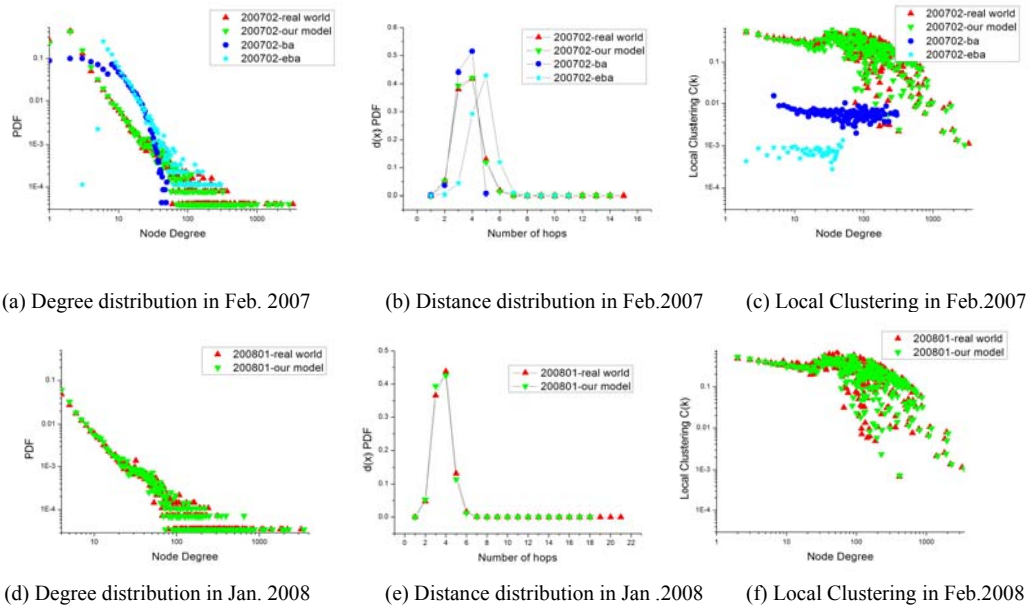

(a) Degree distribution in Feb. 2007  (b) Distance distribution in Feb.2007  (c) Local Clustering in Feb.2007

(d) Degree distribution in Jan. 2008  (e) Distance distribution in Jan .2008  (f) Local Clustering in Feb.2008

Figure 2. The static graph patterns of output graphs

Through the three steps, the model finish inferring $G_{n+1}$ based on $G_1, G_2, \cdots, G_n$. Furthermore, it can inferring $G_{n+2}$ based on $G_1, G_2, \cdots, G_n$ and $G_{n+1}$ (generated by the model just now). If necessary, we can use the model to infer the time series of graphs in the future.

## VI. EXPERIMENTS

In this section, we conduct a number of experiments to validate our model. The input topology we use is from ucla between January, 2004 and January, 2007. We extract one topology for one month. Using our model, we infer the topology graph series in the next 12 months (from February, 2007 to January, 2008), and 12 output graphs totally. All the parameters ($\Delta N$, $a_n$, $a_e$ and $p$) are calculated from the corresponding real-world topology.

First, we use BA and EBA model to generate the graph in February, 2007 too. Looking at Figure 2 (a), (b) and (c), we found that the output graph of our model is closer to real-world graph than the other two models. It matches the degree distribution, distance distribution, and local clustering with the real-world graph very well.

Then, we compare the output graph in January, 2008 in Figure 2 (d), (e) and (f). We found that with the increase of time, the output graphs of our model are close to the real-world graph too. The degree distribution, distance distribution, and local clustering are fitted well too.

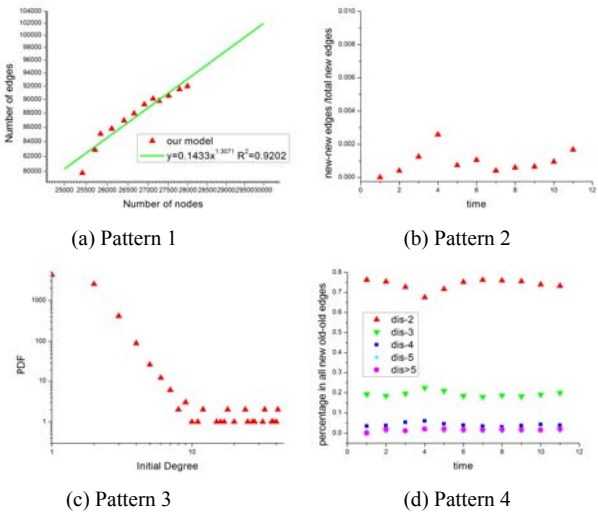

(a) Pattern 1  (b) Pattern 2
(c) Pattern 3  (d) Pattern 4

Figure 3. The dynamic graph patterns of output graphs

Finally, we would like to validate whether the time series of graphs that our model generates can follow dynamic graph patterns. Looking at Figure 3, the four dynamic graph patterns we introduce in section 4 are also found.

From the experiments, we see the output graphs of our model can match both static and dynamic graph patterns very well.

## VII. DISCUSSION

$\Delta N$ shows the increase of graph size. How about the other three parameters? When they change, what will happen in the output graph? Which topology characteristic are they related to? In this section, we answer these questions with theoretical and experimental methods.

### A. Stable factor of nodes $a_n$,

If there are more common nodes between $G_{t+1}$ and $G_t$, the value of $\dfrac{|V_{t+1} \cap V_t|}{|V_{t+1} \cup V_t|}$ is closer to 1. Conversely, it is close to 0. So it can reveal the similarity between $G_{t+1}$ and $G_t$ (or the nodes stability at time $t+1$). We call it the stability coefficient of nodes in this paper.

**THEOREM 1**. In our model, if the increase of graph size $\Delta N$ is constant at every time point, the stability coefficient of nodes at $t+1$ is expected to:

$$\frac{|V_{t+1} \cap V_t|}{|V_{t+1} \cup V_t|} = \frac{(t-1)a_n \Delta N + |V_1| - \Delta N}{t a_n \Delta N + |V_1| + \Delta N} \quad (12)$$

**PROOF**. From the initialization step described in Section 4, we can get

$$|V_{t+1}| = \sum_{i=1}^{t} a_n (1-a_n)^{t-k} |V_i| + \Delta N$$

$$|V_{t+1}| = a_n |V_t| + (1-a_n)(|V_t| - \Delta N) + \Delta N = |V_t| + a\Delta N$$

$$|V_t| = |V_1| + (t-1)a_n \Delta N \quad (13)$$

We can also get the number of elements in the intersection between $V_{t+1}$ and $V_t$

$$|V_{t+1} \cap V_t| = \sum_{i=1}^{t} a_n (1-a_n)^{t-k} |V_i| = |V_1| + (t-1)a_n \Delta N - \Delta N \quad (14)$$

Using (13) and (14), the stability coefficient of nodes at $t+1$ is:

$$\frac{|V_{t+1} \cap V_t|}{|V_{t+1} \cup V_t|} = \frac{|V_{t+1} \cap V_t|}{|V_{t+1}| + |V_t| - |V_{t+1} \cap V_t|} = \frac{(t-1)a_n \Delta N + |V_1| - \Delta N}{t a_n \Delta N + |V_1| + \Delta N}$$

We keep the input time series of graphs and other parameters unchanged and generate graphs with different $a_n$ ($a_n$=0.1, 0.2…, 0.9). Looking at Figure 4(a), the plot matches (12) approximately. Furthermore, with the increase of $a_n$, the stability coefficient of nodes becomes greater. We will do similar experiments for the other two parameters $a_e$ and $p$.

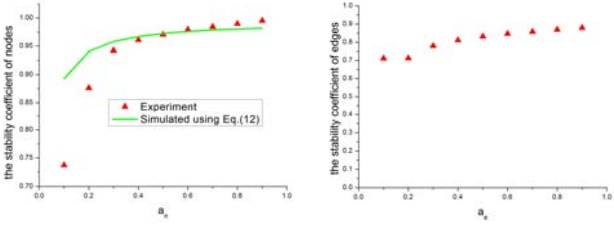

(a) Stable factor of nodes $a_n$    (b) Stable factor of edges $a_e$

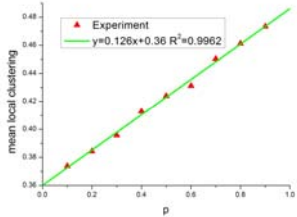

(c) Clustering factor $p$

Figure 4. The parameters used in our model

### B. Stable factor of edges $a_e$

Similarly, we call $\frac{|E_{t+1} \cap E_t|}{|E_{t+1} \cup E_t|}$ the stability coefficient of edges at $t+1$.

**THEOREM 2**. In our model, the stability coefficient of edges at $t+1$ is expected to:

$$\frac{|E_{t+1} \cap E_t|}{|E_{t+1} \cup E_t|} = \frac{\sum_{i=1}^{t} a_e(1-a_e)|V_i|^\alpha}{|V_{t+1}|^\alpha + |V_t|^\alpha - \sum_{i=1}^{t} a_e(1-a_e)|V_i|^\alpha} \quad (15)$$

It can be proved like THEOREM 1. Looking at Figure 4(b), with the increase of $a_e$, the stability of edges gets stronger.

### C. Clustering factor $p$

Now we discuss the cluster factor $p$.

**THEOREM 3**. In our model, the mean local clustering in the output graph grows linearly with the cluster factor $p$.

**PROOF.** If an old node with a degree $k$ is selected as an end-point of a new edge. Its expected local clustering $\hat{C}(k)$ at $t+1$ is

$$\hat{C}(k) = p\frac{C(k)k(k-1)+2}{k(k+1)} + (1-p)\frac{C(k)k(k-1)}{k(k+1)} = \frac{C(k)k(k-1)+2p}{k(k+1)} \quad (16)$$

where $C(k)$ is its local clustering at $t$.

We can easily get the expected local clustering of a new node with an initial degree $k$.

$$\tilde{C}(k) = \frac{pk}{k(k-1)} = \frac{p}{k-1} \quad (17)$$

If assuming all the nodes were selected with same probability, we can get the expected mean local clustering in the output graph of our model.

$$\bar{C} = \frac{(2E+\Delta N)\int \hat{C}(k)P(k)dk + (|V_t|-2E-\Delta N)\int C(k)P(k)dk + \Delta N\int \tilde{C}(k)\tilde{P}(k)dk}{|V_{t+1}|}$$

where E is the number of new old-old edges, $P(k)$ is the degree distribution of $G_t$ and $\tilde{P}(k)$ is the initial degree distribution at $t+1$.

Using (16) and (17), the equation can be simplified to

$$\bar{C} = Ap + B \quad (18)$$

where

$$A = \frac{(2E+\Delta N)\int \frac{2}{k(k+1)}P(k)dk + \Delta N\int \frac{1}{k-1}\tilde{P}(k)dk}{|V_{t+1}|}$$

$$B = \frac{(2E+\Delta N)\int \frac{C(k)(k-1)}{k+1}P(k)dk + (|V_t|-2E-\Delta N)\int C(k)P(k)dk}{|V_{t+1}|}$$

If we keep the input time series of graphs and other parameters unchanged, A and B remain constant. The mean local clustering $\bar{C}$ grows linearly with $p$. In Figure 4(c), we find the results of experiments can fit this perfectly.

## VIII. RELATED WORK

Although quite a lot of studies are about Internet topology analysis and modeling, most of them focus on searching for static graph patterns. Only a few are focusing on dynamic graph patterns. The authors in [16] found Densification power law and shrinking diameters, and proposed a Forest Fire Model to match these two graph patterns. Our work is motivated by their work, which some of our ideas in our model are from. However, there are some key differences between ours and theirs. Our model takes time series of graphs as input, and their model takes a single graph as input. The parameters used in our model are also different from their model. In addition, we find the total graph also obeys a power law, and this improves their Densification power law. The latter three graph patterns are first found by us, but our work does not focus on shrinking diameters they found. The authors in [25] also try to characterize the evolution of Internet topology, but they focuses on solving the liveness problem.

## IX. CONCLUSION

In this paper, we study Internet AS-level topology from two different projects. The following four patterns are founded in both of them.

(1) The number of edges grows super-linearly with the number of nodes. Specifically, it follows a power-law pattern. In addition, there are same patterns in total graph.

(2) With a low probability, an edge between two new nodes is added.

(3) The new nodes have different initial degrees. Furthermore, the initial degree distribution follows a power law pattern.

(4) The shorter the distance between two nodes, the higher the probability they will become neighbors at next time point.

We also propose a model. It takes as input the graph series in the past, the number of new nodes $\Delta N$, the stable factor of nodes $a_n$, the stable factor of edges $a_e$, and the clustering factor $p$. After three steps (preparation, initialization and generation), it outputs the topology graph at next time point. With both theoretical and experimental methods, we prove the parameters used in our work are directly related to some important graph characteristics.

Our work is an attempt to develop a model that can match both static and dynamic graph patterns. We will keep an eye on the evolution of Internet topology. The data colleted in the future will validate our model further. In addition, we will try to apply our model in other complex networks.

ACKNOWLEDGMENTS

We thank the researchers in CAIDA and UCLA for sharing their Internet AS-level topology data with us.